\documentclass{aa}
\usepackage{graphicx}
\usepackage{natbib}
\usepackage{amssymb}

\bibpunct{(}{)}{;}{a}{}{,} 

\begin{document}

\newcommand{\rmd}{{\mathrm d}}
\newcommand{\mm}{\mathrm}
\newcommand{\mi}{\mathit}
\newcommand{\mc}{\mathcal}
\newcommand{\dm}{\dot\mc{M}}
\newcommand{\me}{m_\mathrm{e}}
\newcommand{\mpr}{m_\mathrm{p}}
\newcommand{\mk}{}  
\newcommand{\be}{\begin{equation}}
\newcommand{\ee}{\end{equation}}

\title{The transition from a cool disk to an ion supported flow}

\author{H.C.\ Spruit \and B.\ Deufel  }

\offprints{henk@mpa-garching.mpg.de}
\institute{
       Max-Planck-Institut f\"ur Astrophysik, 
       Postfach 1317, D-85741 Garching, Germany
          }

\date{Received / Accepted }

\abstract{
We show that the inner regions of a cool accretion disk in an X-ray binary can 
transform into an advective, ion supported accretion flow (an optically thin ADAF, 
here called ISAF),
through events  involving only the known properties of the Coulomb interaction
in a two-temperature plasma,  standard radiation processes, and viscous heating. 
The optically thin inner edge of the disk is heated to a few 100 keV by the 
strong flux of hot ions from the surrounding hot ISAF. We show that he
resident ions in this `warm' disk are thermally unstable due to internal viscous
heating, and heat up to their virial temperature. The innermost disk
regions thus evaporate and feed the ISAF. These processes are
demonstrated with time dependent calculations of a two-temperature
plasma in vertical hydrostatic equilibrium, including heating by
external ions, internal proton--electron energy exchange, and viscous
heating. The process complements the `coronal' evaporation mechanism
which operates at larger distances from the central object.
  }

\maketitle

\section{Introduction}
\label{sec:intro}

Accreting galactic black holes (BHC) and active galactic nuclei (AGN)
are often observed with two different spectral components: a soft
component which is probably due to a multi--color blackbody from an
optically thick, geometrically thin standard disk \citep[
SSD]{sunyaev73}, and a hard component which is linked to an optically
thin and geometrically thick flow. The hard component (an approximate
power law with high energy cut off at $E_c\approx 100$ keV) is most
likely produced by inverse Compton scattering of soft photons on a hot
thermal plasma \citep{titar80}. 
\cite{shapiro76} showed that accretion can take place in the form of 
a two-temperature plasma, with the properties needed to produce such
Comptonized radiation. Stable, optically thin, two-temperature flows were
studied by \cite{ichi77} and \cite{rees82}.  Extensive theoretical
work on these accretion flows, 
stressing the role of the advection of internal energy
was done by \cite{narayan94,narayan95a,narayan95b}, for a review see
\cite{narayan98}, see also Esin et al. (1997) and references therein. 

The effects of
advection are the same for the optically thin two-temperature flows 
and optically thick, radiation supported flows, and both types of flow
are now customarily called `ADAF's. The distinction between these types, 
which was explicit in the older labels `ion supported' and `radiation supported',
is still needed in many applications, however. For this reason, we propose
here the name `ion supported accretion flow' or ISAF as a 
means of distinguishing the optically thin type of ADAF from the
radiation supported (`RSAF') type.

\begin{figure*}[t]
  \begin{center}
    \includegraphics[width=12cm]{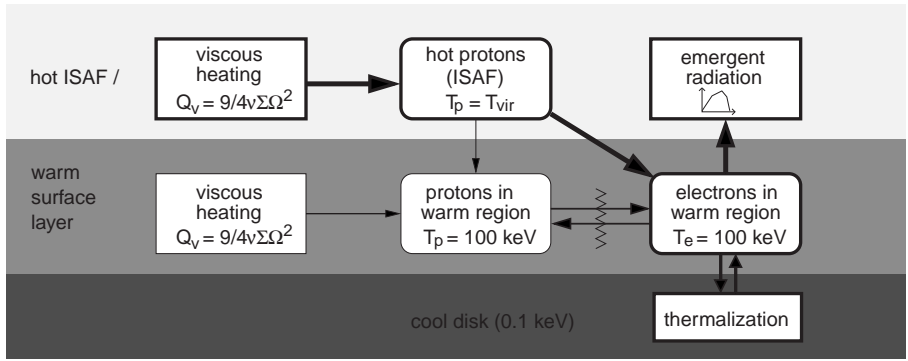} \caption{Energy channels in
      a cool accretion disk embedded in a hot region. The energy
      dissipation in the cool disk is assumed small compared the
      atmosphere (a corona or an ion supported ADAF, here called ISAF). 
      Squared boxes show physical
      processes, round boxes the particles involved.  Heavy arrows and
      boxes show the main energy channel: viscous dissipation in the
      ISAF heats the protons there, which illuminate the cool disk
      below. By Coulomb interactions the ISAF protons loose their
      energy mainly to electrons, producing a warm ($\sim 80$ keV)
      layer which radiates this energy by Compton-upscattering of soft
      photons from the cool disk below.  The protons in the warm
      surface layer are largely {\it outside} this main energy
      channel. Near the transition radius $R_mm{tr}$ the cool part disappears and
      the warm part heats up to several hundred keV.  Then the energy
      exchange of disk protons with the electrons is slow (shown by
      the saw tooth line) and viscous heating of protons becomes
      important. }
    \label{figure}
  \end{center}
\end{figure*}

ISAFs have attracted attention because of their potential 
to explain the spectra of X-ray transients (Esin et al. 1997 and references 
therein). The observations are consistent with an
accretion flow that consists of two zones: an interior ISAF that
extends from the black hole horizon to a transition radius
$R_\mm{tr}$, followed by an optically thick, geometrically thin and
cool standard disk outside $R_\mm{tr}$.  A partial overlap between the
two regions is probable since observations show evidence for the close
vicinity of hot and cold matter in the central regions of BHCs and
AGNs. This is indicated by a K$_\alpha$ iron fluorescence line at
6.4 keV and a Compton reflection component between $\approx10$--30
keV.  

A critical element of such an accretion geometry is the change
from the geometrically thin SSD to the hot ISAF flow at the transition
radius. An alternative interpretation of the spectra is given by the
`magnetic flare' or disk corona model (Haardt \& Maraschi 1991 , 
\nocite{haardt91} \nocite{maraschi97} Maraschi \& Haardt 1997 and 
references therein, di Matteo et al. 1999\nocite{matteo99}, Merloni et al. 2000
and references therein\nocite{merl00}), in which this transition is absent 
and the X-ray emitting plasma is a layer on top of a disk, heated by 
mechanical energy transfer from the disk.

Signatures of a disk transition radius have recently been found
in the power density spectra of Cyg X--1\citep{churazov01}, \citep{gilfanov00}.  
From an analysis of the reflection component
\cite{salvo01} showed that in Cyg X--1 the transition radius $R_\mm{tr}$
is located between $10R_\mm{S}$ and $70R_\mm{S}$, if the
observed reflected spectrum is due to a smeared component, or
$6R_\mm{S}<R_\mm{tr}<20R_\mm{S}$ if the reflection is unsmeared, e.g.
from the companion star or the outer disk.

We take these observations as reasonable indications that a transition
from a cool optically thick to a hot optically thin accretion flow does in
fact take place. But how the SSD--ISAF transition works is still under debate
(cf Manmoto et al. 2000 and references therein\nocite{man00}).
\cite{meyer00} propose that the transition from the cold disk to the
optically thin flow is due to a heat flow by electron conduction from
a hot, friction--heated corona to the cold disk below \citep[see
also][]{meyer94}.  This model has its maximum evaporation
efficiency at a large distance from the hole (a few 100 Schwarzschild
radii).  If, due to a high accretion rate in the cool disk, not all
material is evaporated until that distance, the cool disk will survive
until the last stable orbit. A transition radius further in than 100
Schwarzschild radii is inconsistent with this picture.
\cite{rozanska00} investigate conductive and radiative coupling of an
accretion powered corona with an underlying cool disk.  For low
accretion rates they find that the disk completely evaporates whereas
high accretion rates prevent the SSD--ISAF transition as in
\cite{meyer00}. From a mathematical point of view \cite{abramo98} show
that, if the transition region is not too wide, the region must rotate
with super Keplerian orbital speed. Based on this property
\cite{kato00} demonstrate that trapped low--frequency oscillations are
possible in the transition region.

In this paper we show how the inner disk regions, where the coronal
evaporation process does not work, can evaporate into an ISAF. 
Only a few, well-known ingredients need to be invoked: the coupling
of protons and electrons by the Coulomb interaction in a fully ionized
plasma, standard radiation processes, and viscous heating. 

The starting point is the view, supported by the observations mentioned, 
that an ISAF and a cool disk can coexist. That is, 
there is a partial overlap between the cold disk and the ISAF. In 
the overlap region, there is a very strong interaction between the two, since 
the ISAF consists of energetic ions (10--100 MeV) that penetrate the cool disk 
to a significant depth. The goal in the next sections is to determine the  nature 
of the energy and mass exchange in this interaction region, and to show 
that it will lead to evaporation at the inner edge of the cool disk. The argument 
is then closed by determining the conditions under which these processes can 
consistently lead to the coexistence of the disk and the ISAF that was assumed 
at the outset.

\subsection{Interaction between ISAF and cool disk}
The essence of the processes described below is given by the energetic
interaction between ISAF and disk.
In Fig. \ref{figure} we have sketched the main energy channels 
involved in this interaction. The
ultimate source of energy is the release of gravitational binding
energy. We assume here that a significant fraction (at least a few
tens of per cent) of this energy goes into the protons (on account of
their much higher mass compared to the electrons). This viscous energy
release predominantly takes place in the hot region (the ISAF). The protons and
electrons in the ISAF form a two--temperature plasma, where the
protons have temperatures near their virial temperature. The electrons
are much cooler due to their strong interaction with the radiation
field and the slow rate at which they exchange energy with the ISAF
protons.

Due to their low temperature, the conductive energy flux carried from the ISAF
to the disk by the electrons is negligible. This is one of the reasons why the
coronal evaporation mechanism that functions well at larger distances from
the central mass fails for a two-temperature plasma. The energy flux to the cool
disk is carried almost entirely by the ions. This ion energy flux can not take over
the  role played by electron conduction in the coronal evaporation process, for two 
reasons. One is that the mean free path of the ions is not negligible as it is in the 
case of electrons. The stopping length of  10--100 MeV ions penetrating into a cool
disk corresponds to a Thomson optical depth of order unity. This prevents the
development of the thin energy deposition layer that is needed to heat the plasma
to high temperatures. More important is the fact that the energy carried by these
ions primarily heats the electrons of the disk. These in turn radiate it very
effectively by inverse Compton scattering and limit the plasma temperature in the
interaction region to values of $\sim 100$ keV, well below the virial temperature.
Instead of evaporation, the loss of the ISAF ions to the cool disk is a very
effective {\it condensation} process. This condensation is  an important sink to an
ISAF flow generated by  evaporation at larger distances, and has to be overcome
by a sufficiently powerful evaporation process at some location in the disk. This
location will turn out to be a region near the inner edge of the disk.

\subsection{A `warm' surface layer on the cool disk}
\label{warm}
The ISAF and the cool disk are separated by a `warm' surface skin of
temperature $\sim 100$ keV, intermediate between the ion temperature of the
hot ISAF and that of the cool disk. This heated surface layer is produced by the
energy flux from the ISAF to the cool disk. This flux can  be in the form of
radiation (photons) or particles (ions). Radiative coupling was investigated by
\cite{haardt91,haardt93} in the context of their `two--phase model' for the hard X-ray
spectra of accreting black holes.  Explanation of these spectra as resulting from the
coupling between an ISAF and a cool disk via hot the protons was proposed by
\cite{spruit97} and \cite{spruit00}, and studied in greater detail by \cite{deufel00} 
[\,henceforth paper I\,] and \cite{deufel01a}  [\,henceforth paper II\,].  The
physics of this process has been studied before in the context of accretion onto a
neutron star surface by \cite{zel,alme}, \cite{deufel01} . 

The surface layer radiates
its energy by Compton-upscattering of soft photons from the cool disk below. 
The energy of the ISAF is thus carried to the disk by the hot ISAF protons,  and
radiated away by the disk electrons. The protons in the warm surface layer are
largely outside this main energy channel.

In the present analysis we concentrate on ion heating, which is necessarily
very strong in the overlap region between an ion supported flow and a cool disk.
A radiative flux from the ISAF can exist additionally, but in order to separate
these contributions we assume here that the ISAF itself is radiatively inefficient, so
that its direct contribution to the radiation from the accretion flow negligible.
Finally, the viscous energy release in the warm layer turns out to be small
compared with incident energy flux from the ISAF. Viscous heating  is thus
energetically unimportant in the surface layer, but it will turn out to be crucial to
the evaporation process.

Finally there is the cool disk ($\sim 1$ keV) below the warm layer.
This region is outside of the reach of the impinging hot protons.  The
cool disk serves as a thermalizer for the downward directed radiation
of the warm layer. It is the source of the soft photons which are
Comptonized in the warm region and keep it at moderate temperatures.
The viscous energy release in the cool disk is assumed to be negligible
as a source of soft photons.

In paper II we have shown that close to the inner edge of an accretion
disk, where the surface density (and the optical depth) of the disk
gets small, the penetration of virialized protons heats the entire vertical
disk structure to temperatures of several 100 keV (which is
equivalent to a disappearance of the cold part in Fig.  \ref{figure}).
This corresponds to the first step in our model and is at the same
time the starting point of the present investigation. At high
temperatures the time scale for establishing thermal equilibrium
between the disk protons and electrons is not short compared to the
dynamical time scale any more.  The viscous energy channel (due to
internal heating of protons by friction) gets important now because
coupling to the electrons is weak.

It will turn out that the physics described depends almost only on the
dimensionless radius from the hole $r=R/R_{\rm S}$, the dimensionless
accretion rate $\dot M/\dot M_{\rm Edd}$, and the dimensionless viscosity
parameter $\alpha$. The results are thus scaleable between AGN and BHC 
cases. Where explicit values of the physical parameters are needed, we
take those of a typical BHC case.


\subsection{Coulomb interactions in an ionized plasma}

\label{coul}
In the next three subsections, we briefly review the physics associated with 
the penetration of the protons into the disk. This has been discussed before 
in detail in the references given in the previous subsection. Here, we address 
a few conceptual issues, such as the validity of the approximation that the energy 
of the incident protons is transferred mainly to the electrons in the disk, the accuracy 
of a nonrelativistic treatment, and the charge balance between ISAF and disk.

At the typical energies of the protons incident on the cool disk, the energy 
loss is mostly by long-range Coulomb interactions with the electrons in the 
disk (small-angle scattering on the large number of electrons in a Debye 
sphere). This is opposite to the case of protons with a temperature near 
that of the plasma in which they move. In the latter case, the equilibration 
among the protons is faster than between electrons and protons, by a  
factor of order $(m_\mm{p}/m_\mm{e})^{1/2}$. To see how this apparent 
contradiction is resolved, consider the basic result for the energy loss of a 
charged particle moving in a fully ionized, charge-neutral plasma. This was 
derived by Spitzer (1962) (making use of Chandrasekhar's (1942)\nocite{chandra} 
earlier results on dynamical friction). Introduce as a measure of distance in the 
plasma the Thomson optical depth $\tau$, i.e. $\rmd \tau=\sigma_{\mm T}n 
\rmd l$, where $\sigma_{\mm T}=8\pi e^4/(3m_\mm{e}^2c^4)$ is the 
Thomson cross section, $n$ the electron density and $l$ the distance. The 
rate of change of energy $E=\mpr v^2/2$ of a proton moving with velocity 
$v$ in a field of particles with charge $e$ and mass $m_\mm{f}$ (the `field 
particles' in Spitzer's nomenclature) is then given by Spitzer's Eq. 5-15. In 
our notation, this can be written in terms of  the energy loss length 
$\tau_\mm{f}$ for interaction with the field particles $\mm f$,

\be 
\tau_\mm{f}^{-1}={1\over E}({\rmd E\over \rmd\tau})_\mm{f}=
3\ln\Lambda ({\me\over\mpr})^2({c\over v})^4(1+{\mpr\over 
m_\mm{f}})F(x_\mm{f}),\label{spit}
\ee
where $\ln\Lambda$ is the Coulomb logarithm (which is determined by 
the size of the Debye sphere). Here
\be F(x)=\psi(x)-x\psi^\prime (x), \ee
where $\psi$ is the error function and $\psi^\prime$ its derivative, and 
$ x_\mm{f}= v[{m_\mm f}/(2kT)]^{1/2}$ is (up to a numerical factor) 
the ratio of the incident proton's velocity to the thermal velocity of the field 
particles. The limiting forms of $F$ are
\be 
F(x)\rightarrow x^3/3\quad (x\rightarrow 0),\qquad F\rightarrow 1\quad 
(x\rightarrow\infty).
\ee
We can evaluate (\ref{spit}) under the assumption that the field particles 
that are most relevant for the energy loss are the protons or the electrons, 
respectively, and compare the loss lengths. If the incident proton has velocity 
comparable with the thermal velocity of the field protons, we have $x_\mm{p} 
\sim 1$, and $x_\mm{e}=(\me/\mpr)^{1/2}x_\mm{p}\ll 1$. $F(x_\mm{p})$ is 
then of order unity and $F(x_\mm{e})\approx x^3_\mm{e}/3$. Setting 
$\mm{f}=\mm{e}$ respectively $\mm{f=p}$ in (\ref{spit}) and taking the ratio, we have
\be {\tau_\mm{e}\over\tau_\mm{p}}\approx ({\mpr\over\me})^{1/2}. \ee
The loss length for interaction with the electrons is thus much longer than for 
interaction with the protons, and the interaction with electrons can be neglected. 
This is the well known result for the relaxation of a proton 
distribution in a plasma that is not too far from its thermal equilibrium.

For incoming protons of high energy, however, the result is different because 
$x_\mm{e}$ is not sufficiently small any more. In the high-$v$ limit, $F(x_\mm{e})=
F(x_\mm{p})=1$, and one has
\be {\tau_\mm{e}\over\tau_\mm{p}}\approx 2{\me\over\mpr}=10^{-3}. \ee
In this limit, the energy loss is thus predominantly to the electrons. A related 
case is that of the ionization losses of fast particles in neutral matter (for 
references see Ryter et al. 1970). The case of an ionized plasma is simpler, 
since the electrons are not bound in atoms. The change from proton-dominated 
loss to electron-dominated loss takes place at an intermediate velocity $v_{\mm c}$, 
at which $F(x_\mm{p})\approx 1$ but $x_\mm{e}$ still small, so that 
$F(x_\mm{e})\approx x^3_\mm{e}/3$. Equating $\tau_\mm{e}$ and 
$\tau_\mm{p}$ then yields
\be x_{\mm e,c}\approx ({\me\over\mpr})^{1/3},\ee
or
\be {E_{\mm c}\over kT}\approx ({\mpr\over\me})^{1/3}\approx 12.\ee
For the electron temperatures we encounter in our models, $T\sim 100$ keV, 
energy loss to the field protons can thus be neglected for incoming protons 
with energy $E\ga 1$ MeV. This is the case in all calculations presented here.

\subsection{Corrections at high and low energies}

Spitzer's treatment is non-relativistic, while virialized ISAF protons near 
the hole can reach sub-relativistic temperatures. A fully relativistic treatment 
of the Coulomb interactions in a plasma has been given by Stepney \& Guilbert (1983). 
We have compared the classical treatment according to Spitzer's theory with 
this relativistic result in Deufel et al. (2001), and found it to be accurate to better 
than 5\% for proton temperatures $<100$ MeV. The classical approximation 
in Spitzer's analysis therefore does not introduce a significant error for the 
problem considered here. 

For high energies, the Coulomb energy loss becomes so small that loss by 
direct nuclear collisions becomes competitive. This happens (cf.\ Stepney \& 
Guilbert 1983) at $E\ga 300$ MeV, an energy that is not reached by virialized 
protons except in the tail of their distribution. We ignore these direct nuclear 
collisions. Note, however, that a gradual nuclear processing by such collisions 
can be important (Aharonian \& Sunyaev, 1984)\nocite{aha}, in particular for the 
production of the Lithium. The Lithium overabundances seen in the companions 
of LMXB (Mart\'{\i}n at al., 1994a)\nocite{martin94a}, may in fact be a 
characteristic signature of  the interaction of an ADAF and a disk described 
here (Mart\'{\i}n et al. 1994b, Spruit 1997)\nocite{martin94b}\nocite{spruit97}.

As the protons slow down, they eventually equilibrate with the field protons. 
This last part of the process is not accurately described by the energy loss 
formula (\ref{spit}). In addition to the simple energy loss of a particle moving 
on a straight path through the plasma, one has to take into account the random 
drift in direction and energy resulting from the interaction with the fluctuating 
electric field in the plasma. This drift can be ignored to first order (end of Sect. 
5.2 in Spitzer 1962), but takes over in the final process of equilibration with the 
plasma. This last phase involves negligible energy transfer compared with the 
initial energy of the protons in the present calculations, and can be ignored here.

\subsection{Charge balance}

The protons penetrating into the disk imply a current that has to be balanced 
by a `return current'. As in all such situations, this return current results from 
the electric field that builds up due to the proton current. As this electric field 
develops, it drives a flow of electrons from the ADAF to the disk which maintains 
the charge balance. Since the electron density in the disk is high, the return current 
does not involve a high field strength.

\section{The evaporation process}
\label{sec:mechanism}

Any mechanism heating the protons in the disk, even if only small, can 
potentially give rise to a thermal instability since the disk protons do not loose 
energy efficiently. Their energy loss by radiation is negligible on account of their 
high mass, and the transfer of energy to the disk electrons by Coulomb interaction 
is inefficient 
in a sufficiently hot plasma. In addition, by the nature of the Coulomb interaction, 
the time scale for the disk protons to equilibrate with the disk electrons {\it increases} 
with temperature (cf Sect.\ \ref{coul}). Depending on the strength of the 
mechanism heating the protons and its temperature dependence, a runaway may 
occur in which the disk protons continue to heat up as their cooling rate continues 
to decrease. 

One heating process to be taken into account is 
internal viscous energy release (due to friction). At the low surface densities in 
the region of interest, the amount of energy released by friction is small compared 
with the heating flux by the external ISAF protons. If we may assume that a 
reasonable fraction of the viscous energy release goes into the protons, this small 
amount can still be quite important for the energy balance of the disk protons because 
of their low energy loss rate, and must therefore be taken into account. A second 
heating process is energy transfer from the incident ISAF protons to protons in the disk
by Coulomb  interaction. Though we have seen ( Sect.\ \ref{coul}) that the incident protons 
loose their energy mainly to the electrons, the small amount transferred to the disk 
protons may conceivably be relevant for their energy balance. 
In the following we investigate these heating mechanisms in turn, an conclude that 
viscous heating can give rise to a runaway, while heating of the disk protons by Coulomb 
interaction with the incident protons does not.

\subsection{Viscous heating of the disk protons, instability}
\label{sec:vis}

The energy balance of the protons in the warm disk is conveniently
described in terms of heating and cooling time scales due to the processes
involved. First we consider the cooling time scale by transfer of energy from protons 
to electrons in the disk. 

The energy exchange timescale by Coulomb interactions between nonrelativistic 
protons and electrons  with Maxwellian distributions at temperatures 
$T_\mm{p}$ and $T_\mm{e}$, respectively, is given by Spitzer's (1962)  
classical result:

\begin{equation}
t_\mm{ep}=\frac{3\,m_\mm{p}\,k^{3/2} (T_\mm{e} +
        \frac{m_\mm{e}}{m_\mm{p}}T_\mm{p})^{3/2}}
        {8(2\pi\,m_\mm{e})^{1/2}\,e^4 \,n_\mm{e} \ln\Lambda} \simeq 253\,
        \frac{T_\mm{e}^{3/2}}
        {n_\mm{e} \ln\Lambda}\;\;\mm{s},
\label{eq:tsclep}
\end{equation}
where $m_\mm{p}$,\,$m_\mm{e}$ are the mass of the proton and electron,
respectively, $k$ is the Boltzmann constant, $e$ the charge of the
electron, $n_\mm{e}$ the electron density and $\ln \Lambda \approx 20$
the Coulomb logarithm. The approximate equality is used since we
consider cases where initially the electron temperature is of the
order of the proton temperature and therefore the contribution of
$\frac{m_\mm{e}}{m_\mm{p}} T_\mm{p}$ can safely be neglected.  

The relativistic correction to this result is small for the conditions of
interest here. In \cite{deufel01a} we have shown that Spitzer's formalism 
is accurate to better than 5\% compared to the relativistic treatment by 
\cite{stepg83}, for proton energies below 100 MeV and
electron temperatures $kT_\mm{e}\lesssim 50$ keV.

Next, the viscous heating timescale in an accretion disk is given by

\begin{equation}
t_\mm{th}=\frac{1}{\alpha \Omega}\;,     
\label{eq:tsclvis}
\end{equation}
where $\Omega = (G M / R^3)^{1/2}$ is the Kepler angular velocity and
$\alpha$ is the viscosity parameter \citep{sunyaev73}.

Comparing Eqs. (\ref{eq:tsclep},\ref{eq:tsclvis}) yields a 
critical electron density

\begin{equation}
n_\mm{e}^{*}= 
\frac{3\,m_\mm{p}k^{3/2}}{8\sqrt{2\pi\,m_\mm{e}}\,e^4\,\ln\Lambda}\,T_\mm{e}^{3/2}
\,\alpha\,\Omega\;.
\label{eq:nstar}
\end{equation}

For electron densities $n_\mm{e}<n_\mm{e}^{*}$ viscous heating in an
accretion disk works faster than proton--electron coupling. In
those regions the protons can not loose their energy fast enough, and
heat up. This increases $n_{\rm e}^*$. At the same time, the higher
proton temperature causes the layer to expand, by hydrostatic
equilibrium.  On both accounts, the Coulomb coupling between protons
and electrons decreases, and the heating of the protons accelerates.
Thus we expect viscous heating of the protons to lead to thermal instability 
wherever the density is less than given by Eq.~(\ref{eq:nstar}). Since the
density decreases steeply with height in the atmosphere, there is
always a level above which it is unstable. A new equilibrium is
reached only when the protons reach the virial temperature. The
unstable part of the atmosphere has then expanded to $H/R\sim 1$, and
it is then, effectively, part of the ISAF in which the disk is
embedded. The unstable part of the atmosphere has {\it evaporated},
feeding the ISAF.  Whether such an instability has an effect on the
global accretion properties depends on the mass in the unstable
region.

The amount of mass involved in the instability is the mass in the
atmosphere above the level where the density has dropped to $n_{\rm
  e}^*$. Let this mass be ${\mc M}^*$ (per unit surface area of the disk).
The time scale on which the protons heat up in the unstable region is
just the viscous heating time scale $t_{\rm th}$. The rate at which
the atmosphere evaporates is thus approximately

\begin{equation}
\dot\mc M \approx \alpha\Omega \mc M^* \;.
\label{eq:evap1}
\end{equation}

The unstable mass $\mc M^*$ is found by integrating the density in the
atmosphere upward from the level $z^*$ where $n=n^*$:

\begin{eqnarray}
    \mc{M}^*  =  \int_{z^*}^\infty m_\mm{p}\,n_\mm{e}(z)\,\rmd z 
     =  m_\mm{p}\;n_\mm{e}^*\; \,H \sqrt{2}\, f(u^*) \;\;,
    \label{eq:mstar}      
\end{eqnarray}
where $H$ is the scale height of the disk atmosphere (assumed
isothermal) with density stratification,
$n_\mm{e} =n_0 \exp[-u^{2}]$ with $u^2 = z^2/2H^2$. The function
$f(u^\star)$ is

\begin{eqnarray}
  f(u^\star)  =  \int_{u^\star}^{\infty}\exp[u^{\star2}-u^2]\,\rmd u 
   =  
  \frac{\sqrt{\pi}}{2}\, e^{u^{\star2}}\,\left(1-\mm{erf}[u^{\star}]\right)\;\;.
  \label{eq:f}
\end{eqnarray}

To evaluate Eq.~(\ref{eq:f}) we need to know the dimensionless
critical height above the midplane, $u^\star$.  The level at which
$n_{\rm e}=n_{\rm e}^*$ depends on the surface density of the disk,
being higher at large surface density. We measure the surface density
by the Thomson scattering optical depth $\tau_{1/2}$, measured from the
midplane to the surface of the disk. In terms of the density at the
midplane $n_0$, this is:

\begin{equation}
\tau_{1/2} = n_0 \, \sigma_\mm{T} H\,\sqrt{\pi/2}\;.
\end{equation}

With Eq.~(\ref{eq:nstar}) we can then calculate $u^{\star}$:

\begin{equation}
u^{\star}=\left[-\ln\left(\frac{\pi\,m_\mm{p}^{1/2}\,k^2}
{\sqrt{2}\,m_\mm{e}^{5/2}\,c^4\,\ln\Lambda}\frac{\alpha
  \,T_\mm{e}^2}{\tau_{1/2}}\right)\right]^{1/2}\;\;.
  \label{eq:ustar}
\end{equation}

Above a certain temperature (for a given $\alpha$) $u^*$ is not
defined, as the value within the square brackets in
Eq.~(\ref{eq:ustar}) drops below zero. This means that
the \em whole \rm atmosphere is subject to the instability, and we
can set $u^\star=0$.  In this cases $f(u^\star)=\sqrt{\pi}/2$ (the
maximum value for $f$).

Using Eq.~(\ref{eq:mstar}) to replace $\mc{M}^\star$ in Eq.~(\ref{eq:evap1}) 
this yields, with Eq.~(\ref{eq:nstar}):

\begin{eqnarray}
\dot\mc{M} = 
\frac{3}{4\sqrt{2\pi}}\frac{m_p^{3/2}\, k^2}{m_e^{1/2}\,e^4\,\ln\Lambda}
\,\alpha^2\,\Omega\,T_\mm{e}^2\,f(u^\star)\;\;.
\label{eq:evap}
\end{eqnarray}
Here $H=c_s/\Omega$ and $c_s=\sqrt{2\,k\,T/m_\mm{p}}$ is the
isothermal sound speed.

An evaporating part exists in every disk atmosphere, but at low
temperatures only the highest layers evaporate. The mass loss is
unimportant in such a case and can not change the properties of
the disk. Eq. (\ref{eq:evap}) shows that the evaporation rate is
a relatively strong function of $R$, $\alpha$ and $T_\mm{e}$.
Whether situations exist where this evaporation rate becomes important is 
investigated in Sect. \ref{sec:visvscon}.

\subsection{Heating by proton-proton interactions}
\label{sec:pp}

In this subsection we show that Coulomb interaction of the incident
protons with the resident disk protons is relatively unimportant for
the conditions encountered. It is included in the numerical results reported
in the next section, however.

The main energy channel is from the hot protons to the electrons in
the warm surface layer (see  Sect.\  \ref{coul} and paper II). The properties of
the Coulomb interactions at the temperature of the incoming protons
are such that only a small fraction of their energy is transferred to
the protons in the layer (which we will call `field protons' here
following the terminology in Spitzer 1962). The energy budget of the
field protons is also small, however (cf. Fig.~\ref{figure}), so the effect
of this channel on the proton temperature in the warm layer has
to be considered.

The interaction of a fast proton with the protons in a much cooler plasma
is given by Spitzer's result discussed in  Sect.\ \ref{spit}. In terms of the rate 
of change of the energy of the incoming proton, it can be expressed as

\begin{equation}
\frac{\rmd E_\mm{p^{'}}}{\rmd t}=-\frac{8\pi\,e^4}{m_\mm{p}
  v_\mm{p'}}\,n_\mm{p}\,\ln\Lambda\,[\psi(x)-x\psi'(x)]\;,
\label{eq:depro}
\end{equation}
where $n_\mm{p}$ is the density of the (cool) field protons. The
subscript $\mm{p}'$ indicates the hot penetrating protons and $\mm{p}$ the
field protons.  Here $\psi(x)$ and $\psi'(x)$ are the error function
and its derivative and $x^2=m_\mm{p}v_\mm{p}^2/2kT_\mm{þ}$ is the
ratio of the velocity of the incoming protons to the thermal velocity
of the field protons. In the following we set $\psi-x\psi'=1$. This
approximation is valid if the velocity of the incoming protons much
exceeds the velocity of the field protons. In our case we consider
virialized protons penetrating into the considerably cooler disk
plasma, so the approximation is valid.

The heating rate $W_\mm{p'\!p}$ per unit volume due to the interaction
of hot protons with the field protons is then

\begin{equation}
  W_\mm{p'\! p} = n_\mm{p'}\cdot\frac{\rmd E_\mm{p'}}{\rmd t}
  = \frac{8\pi\,e^4}{m_\mm{p} v_\mm{p'}^2}\,\ln\Lambda\,n_\mm{p}
  \,n_\mm{p'}\,v_\mm{p'}\;\,.
\label{eq:pprate}
\end{equation}

This can be compared with the rate of energy transfer from the field protons to
the electrons in the warm layer as given by
Eq. (5-30) from \cite{spitzer62},

\begin{eqnarray}
  W_\mm{pe} &  = &  -\,n_\mm{e}\,k\,\frac{T_\mm{e}-T_\mm{p}}{t_\mm{ep}} 
  \nonumber \\
  &  \simeq &
  - \frac{8(2\pi\,m_\mm{e})^{1/2}\,e^4\, n_\mm{e}^2
  \ln\Lambda}{3\,m_\mm{p}} 
   \frac{kT_\mm{e}-kT_\mm{p}}{(kT_\mm{e})^{3/2}}\;,
\label{eq:perate}
\end{eqnarray}
where we have used Eq.~(\ref{eq:tsclep}) to replace $t_\mm{ep}$. The
approximate equality indicates that the contribution from
${m_\mm{e}}/{m_\mm{p}}T_\mm{p}$ can again be neglected. Using
$v_\mm{p'}=\sqrt{2kT_\mm{p'}/m_\mm{p}}$, we compare
Eqs.~(\ref{eq:pprate},\ref{eq:perate}) and obtain an estimate for the
equilibrium temperature of the disk protons exposed to penetrating hot
protons:

\begin{eqnarray}
T_\mm{p} \approx
  T_\mm{e} + 114 
    \left(\frac{n_\mm{p'}}{n_\mm{e}} \right)
    \left(\frac{T_\mm{e}}{T_\mm{p'}} \right)^{\!\frac{1}{2}} \,T_\mm{e}\;.
  \label{eq:tpro}
\end{eqnarray}

If the mass flux in the ISAF and its viscosity parameter $\alpha$
are of the same order as in the cool disk, a thin disk approximation for 
both flows yields 
$n_\mm{p'}/n_\mm{e}\sim  (T_\mm{e}/T_\mm{p'})^{3/2}$.
In the inner region of the accretion flow, where $T_\mm{e}\sim 100$ keV and
$T_\mm{p'}>10$MeV, the second term in (\ref{eq:tpro}) is then negligible.

We conclude that heating of the ions in the warm disk by the 
incident ions alone is not sufficient to drive the ion temperature
away from equilibrium with the electrons. For this to happen,
viscous heating has to be included.

\subsection{Numerical simulation of the evaporation} 

We test the inferred heating instability numerically, 
starting with a warm disk of moderate optical depth and
investigating its temporal evolution due to internal viscous heating, and
p$'$--p and p--e energy exchange as described in the previous subsection.

We use a plane--parallel, one--dimensional model geometry. The
vertical density distribution through the atmosphere is found from the
equation of hydrostatic equilibrium as in paper II, except that here we do
not account for the pressure exerted by the decelerating protons.

The temporal evolution of the temperature profile is computed with
a fourth order Runge--Kutta method.  The timestep $\Delta t$ of each
integration is set to the shortest timescale of the heating and cooling 
processes involved. The change of temperature per timestep within a
certain volume is due to the rate of change of the enthalpy there,
$\Delta_t w = \rho c_p \Delta_t T$.

The temperature increase per timestep as a function of optical depth
due to viscous heating of the protons is then given by

\begin{equation}
        \frac{\Delta T_\mm{visc}(\tau)}{\Delta t}= \frac{9}{4}\,\alpha\,
        n_\mm{e}(\tau)\, k \, T_\mm{p}(\tau)\, \Omega_\mm{K}
        \cdot \frac{1}{\rho(\tau)\,c_p}\;,
        \label{eq:dtvisc}
\end{equation}
where $c_p$ is the specific heat at constant pressure.

 The temperature change of the field protons due to the impinging hot
protons is computed similar to the proton illumination method
described in detail in paper I and II. Thus we follow protons from a
Maxwellian distribution with virial temperature through the warm atmosphere
of the disk and record their energy losses. The energy flux in the illuminating
protons is related to the mass condensation rate
[see Eq.~(\ref{eq:massflux}) and Sect. \ref{sec:visvscon} for details]. 
The kinetic energy decrease of the fast protons (p$'$) due to their
interactions with the field protons (p) as a function of
optical depth can be derived from Eq.~(\ref{eq:depro}). This gives the
heating rate $\Lambda_\mm{p'p}$ and the contribution of p$'$p exchange 
to the temperature change per time step is

\begin{equation}
        \frac{\Delta T_\mm{p'p}(\tau)}{\Delta t}=
        \frac{\Lambda_\mm{p'p}}{\rho(\tau)\,c_p}\;.
\label{eq:dtpp}
\end{equation}

Finally the rate of change of the temperature due to the
electron--proton [ep] coupling is [according to \cite{spitzer62}]

\begin{equation}
        \frac{\Delta T_\mm{ep}(\tau)}{\Delta t}=
        \frac{T_\mm{e}(\tau)-T_\mm{p}(\tau)}{t_\mm{ep}(\tau)}\;.
        \label{eq:dtep}
\end{equation}

\begin{figure}
  \includegraphics[width = \hsize]{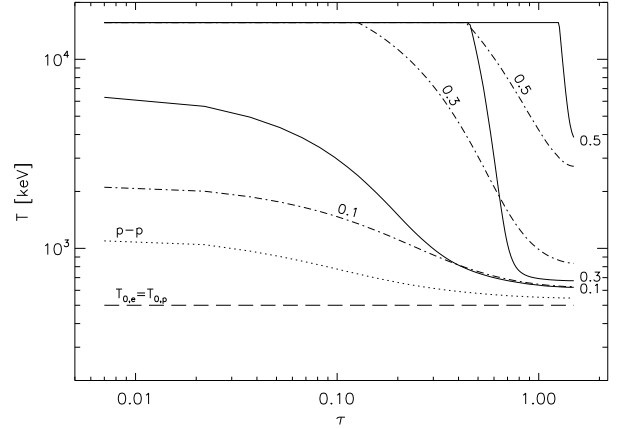}
  \caption{Evolution of the temperature profile due to the viscous
    instability of the protons. The model layer has optical depth
    $\tau_{1/2}=1.5$ and initial temperature $T_\mm{0,p} = T_\mm{0,e}
    = 500$ keV (dashed line).  Dashed--dotted  lines show the profile after 
    10 thermal time scales, solid lines after  t=20. Values of the viscosity
    parameter $\alpha$ are 0.1, 0.3 and 0.5 (numbers
    at the lines).  For comparison the temperature profile after
    $t=20\,t_\mm{th}$ due to $\mm{p'p}$ interactions alone is also
    shown (dotted line).}
  \label{fig:evprof}
\end{figure}

To complete the model the electron temperature in the disk 
needs to be specified, which is regulated by the radiation
processes. We have considered these in some detail in paper II, where
we found that for the present conditions electron temperatures
of 100-500 keV result. The radiation losses by inverse Compton scattering 
increase quite rapidly with $T_\mm{e}$  and with the increasing
optical depth when pair production sets in. This makes the electron 
temperature relatively insensitive. For the present purpose,
it is sufficient to keep the electrons at a constant temperature, which we treat
as a parameter of the model.

When the disk protons in our simulation have reached their local virial
temperature, we do not allow a further temperature increase. In
a more realistic calculation, this limit on the temperature would come
about through the advection of internal energy. Such a calculation
requires a more detailed model of the ISAF and its sources and
sinks of mass and energy, which is beyond the scope of the present
calculation.

As an example we consider a warm disk with an optical
depth (from the surface to the midplane) $\tau_\mm{1/2}=1.5$, an
initially isothermal temperature profile $T_\mm{e}=T_\mm{p}=500$ keV,
illuminated by an ISAF with accretion rate $\dot M_\mm{I}=0.1
\dot M_\mm{Edd}$, where $\dot M_\mm{Edd} = 4 \pi c
R_\star m_\mm{p} /\sigma_\mm{Th}$ is the Eddington accretion rate.

\begin{figure*}
  \sidecaption\includegraphics[keepaspectratio=false,width=12cm]{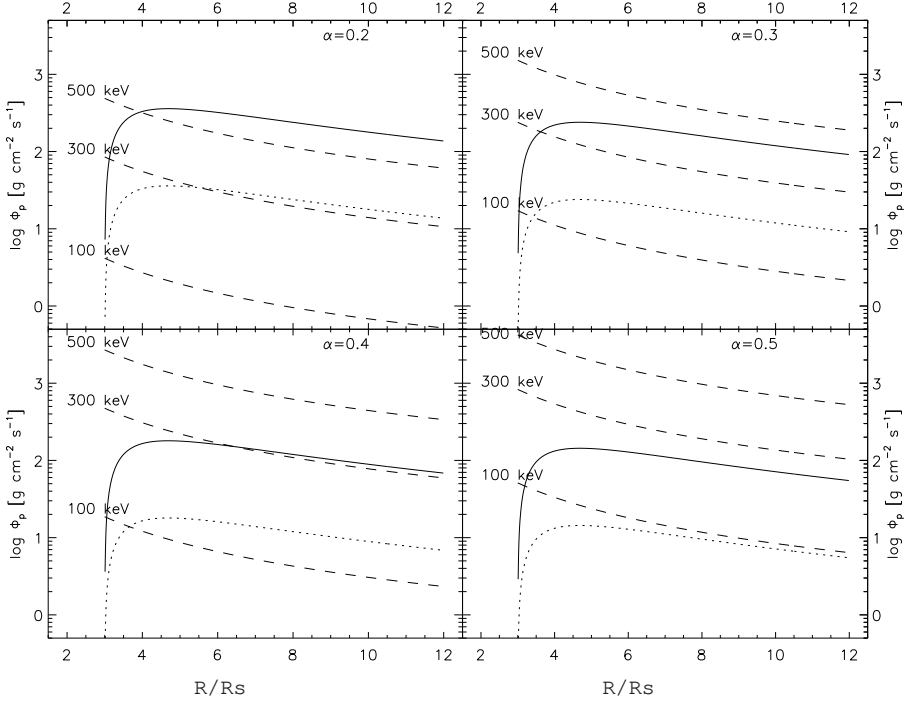}
\caption{Local
  evaporation and condensation rates from a warm (100-500keV) layer of
  optical depth $\tau_\mm{1/2}=1.5$, formed by proton illumination of
  an initially cool disk. Solid line shows the 
  condensation rate onto the layer from an ISAF accreting at $0.1\dot M_{\rm E}$,
  as a function of distance from the hole.  Dashed lines show the
  evaporation rates from the layer, for three values of its
  temperature. In the innermost region, the net effect is evaporation,
  feeding the ISAF, further out the ISAF condenses onto the layer. At
  lower ISAF accretion rates ( $0.01\dot M_{\rm E}$, dotted) the evaporating 
  region becomes wider (see text for details). } \label{fig:evrate}
\end{figure*}

Fig.~\ref{fig:evprof} shows the temperature profiles for various
values of the viscosity parameter $\alpha$ after $t=10\,t_\mm{th}$
(dashed--dotted lines) and $t=20\,t_\mm{th}$ (solid lines), where
$t_\mm{th}=1/(\alpha\Omega)$ is the thermal time scale of
the disk. For $\alpha=0.5$ all disk protons heat up to the virial
temperature, whereas for $\alpha=0.3$ and $\alpha= 0.1$ only a part of
the disk is subject to the instability. The extent of the unstable
part is in excellent agreement with our estimate in Sect.~\ref{sec:vis}. 
In test calculations in which only $\mm{p'p}$ heating and $\mm{ep}$
energy coupling are included, the temperature increase of the disk protons is
considerably smaller than with the viscous heating included. This
confirms our conclusion from Sect.~\ref{sec:pp} that $\mm{p'p}$
interactions alone can not heat the bulk of the internal protons
considerably above the ambient electron temperature [\,cf.
Eq.~(\ref{eq:tpro})].

Summarizing our investigation of the energy channels in a warm disk 
we conclude that viscous energy release in a hot
plasma causes a runaway temperature increase in the unstable upper part
of the disk atmosphere, until the protons have reached their local
virial temperature. The process affects a significant part of the stratification
if the electron temperature is above $\sim 100$ keV.

\section{Evaporation vs. condensation rates}
\label{sec:visvscon}

In Sect. \ref{sec:mechanism} we have shown that the internal
viscous heating of the disk protons leads to a mass evaporation rate
according to Eq.~(\ref{eq:evap}).

At this point we do not yet know  whether the mass loss $\dot\mc{M}$
from a warm disk region is high enough to completely evaporate the disk.
At the same time as the upper atmosphere of the warm disk evaporates, 
the hot protons from ISAF condense into it and increase the surface density. For
an effective evaporation of the disk the mass loss rate must 
be higher than the condensation rate. To compute the condensation rate
we need an estimate of the density in the ISAF.

In our previous numerical simulations of warm disks (paper II)
we parameterized proton mass flux from the ISAF by scaling the
energy flux of the incident protons with the local energy dissipation
rate in the ISAF. Here we adopt a slightly more realistic mass flux rate in an
ISAF.

In a thin disk approximation the surface density $\Sigma_\mm{I}$ of the ISAF,
with accretion rate  $M_\mm{I}$,  is
\begin{equation}
\Sigma_\mm{I}=\frac{\Omega_\mm{K}}{3\pi\alpha c_s^2}\, M_\mm{I} 
        \,[\,1-(3R_\mm{S}/R)^{1/2}]\;.
\label{eq:mdotadaf}
\end{equation}

We assume that the protons in the ISAF have a Maxwellian velocity distribution according to
their local virial temperature. The mass flux rate $\phi$ [in
g\,cm$^{-2}$\,sec$^{-1}$] from a Maxwellian through a surface is given by

\begin{equation}
        \phi=\rho_\mm{p}\,\sqrt{\frac{k\,T_\mm{p}}{2\pi\,m_\mm{p}}}\;.
\label{eq:maxflux}
\end{equation}

From Eq.~(\ref{eq:mdotadaf}) we can calculate the proton
density $\rho_\mm{I}=\Sigma_\mm{I}/(2 H_\mm{I})$ in the ISAF.
This yields the local mass flux rate onto the cool disk for protons at
their virial temperature, i.e. the condensation rate:

\begin{equation}
\phi_\mm{I}=\frac{1}{\alpha}\frac{\dot M_\mm{I}}{8\pi^{3/2}
R^2}\,[\,1-(3R_\mm{S}/R)^{1/2}]\;.
\label{eq:massflux}
\end{equation}

\begin{figure}[t]
\begin{center}
  \includegraphics[width = \hsize]{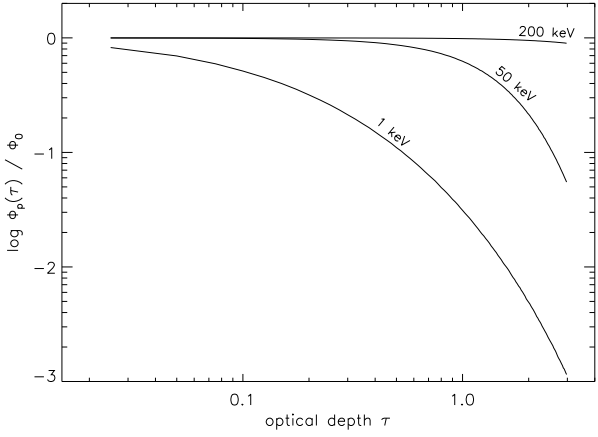} \caption[Decrease of
  the proton mass flux at different temperatures]{Decrease of
    the proton mass flux with depth in a thin, hydrostatic, isothermal disk with
    total optical depth $\tau_\mm{tot}= 2\cdot\tau_\mm{1/2}=3$, at
    disk temperatures of 1, 50 and 200 keV, for
$R=8R_\mm{S}$ and $\alpha=0.1$. At temperatures above 100 keV, only a 
fraction of the incident protons is captured by the disk}  \label{fig:dp}
\end{center}
\end{figure}

We can now compare the mass condensation rate with the evaporation
rate according to Eq.~(\ref{eq:evap}). The evaporation rate is
$\propto (\alpha T)^2$. Therefore high disk temperatures and high
$\alpha$'s favor mass evaporation. In paper II we have calculated 
these temperatures with a radiative transfer calculation that includes
brems photon production, $\gamma\gamma$ pair production and
Comptonization. We found there that the temperature at the
inner edge of a cool accretion disk reaches several 100 keV (`warm state'), 
while its optical depth $\tau_\mm{1/2}$ (at the border with the cool part of 
the disk) is of the order unity for accretion rates around $0.1\dot 
M_\mm{Edd}$. We assume here that at a certain radius the warm disk state
exists and, ignoring its radial structure, compare the local evaporation 
and condensation rates.

Fig.~\ref{fig:evrate} shows this comparison for different values for
the viscosity parameter $\alpha$, different temperatures and
accretion rates.  For values of the viscosity parameter $\alpha\gtrsim
0.3$ and $\dot M=0.1\dot M_\mm{Edd}$ evaporation dominates when 
the temperature of the warm state exceeds $\sim 300$ keV.
As the ISAF density (i.e. the accretion rate) decreases, the condensation 
rate of the protons into the the disk decreases and evaporation dominates 
over the condensation rates over a wider range of radii.  

The condensation rate given by Eq.~(\ref{eq:massflux}) is actually an
overestimate, since it assumes that all incident protons are stopped
in the disk. While this is correct for cool disks, for a high
temperature plasma the rate of the electron--proton energy exchange is
small and a disk with low surface density gets optically thin for the
penetrating hot protons.  This is demonstrated by Fig.~(5) where we
show 
how the incident proton flux changes with depth into a warm disk.  The
Thomson optical depth in this example is ($\tau_\mm{tot} =
2\cdot\tau_{1/2}=3$). At the temperatures of a cool standard disk 
($\la 1$ keV) almost all protons are absorbed in this layer. But at the high 
temperatures of the warm state the disk is optically
thin and practically all ISAF protons fly through the disk {\em without}
being absorbed. The penetrating protons do not add to the surface
density in this case, and evaporation should therefore be possible even at
lower values of $\alpha$. 

At temperatures below 100 keV, the evaporation rate from the warm state 
into the ISAF is quite small and can not balance the loss from it by 
condensation. This is roughly the temperature of the warm surface layer 
on a cool optically thick disk heated by proton illumination (the situation
sketched in Fig 1). Thus the relative importance of evaporation and
condensation reverses just at the point where the cool component
disappears. As long as a cool disk is present, the thermal instability
in its warm surface layer is relatively weak, while it effectively
absorbs all incoming protons. Once the cool component is gone, the
temperature and evaporation rate goes up, while at the same time mass 
condensation by stopping of protons in the disk becomes ineffective.

\subsection{Dependence on the accretion rate}
Figs.~\ref{fig:evrate},\ref{fig:dp} show that net evaporation takes
place close to the hole preferentially at {\it low} accretion rates in
the ISAF, which may sound counterintuitive. It is a
consequence of the fact that the evaporation rate does not depend on
the flux of incident protons (cf.  Eq.~\ref{eq:evap}). This again is a
result of the fact that the field protons in the warm disk lie
`outside the main energy channel': the incident proton energy flux
sets the electron temperature (through the Comptonization balance) but
does not affect the proton temperature directly.
The temperature of the field protons is determined
by the secondary balance between viscous heating of the field protons
and their energy loss to the electrons.

At low accretion rates, however, the hot proton flux eventually
becomes insufficient to keep the layer `warm': it will cool down to
low ($\sim 1$ keV) temperatures by bremsstrahlung losses. We have
studied this transition in paper II, where we found that a layer of
optical depth $\tau_{1/2}=1.5$ can just be kept in the warm state for
an accretion rate in the ISAF of about $0.1\dot M_{\rm E}$. 

At lower accretion rates, a warm disk is possible only when the bremsstrahlung
losses are lower, at lower electron densities. Since the electron
density is proportional to the optical depth $\tau$ of the disk, and the brems
losses
proportional to $n_{\rm e}^2$, the minimum ISAF accretion rate needed
to maintain a warm disk state by proton illumination scales as
$\tau^2$. The optical depth of the disk vanishes towards its
inner edge, so we expect that there is always a region close to the
inner edge of the disk where evaporation takes place, even at very low
accretion rates. 

Summarizing this argument, the optical depth $\tau_\mm{m}$ of a warm disk 
at the point where it matches onto the cool disk,  depends on the accretion rate 
in the ISAF as

\begin{equation}
 \tau_{\rm m}= 2\,\cdot \tau_{{1/2},\mm{m}}\approx 3 \left({\dot
  M_{\rm I}\over 0.1 \dot M_{\rm E}}\right)^{1/2}.
\label{eq:taum}
\end{equation}

\section{An evaporating cool disk}
\label{sec:diskevap}

\subsection{Thin disks with evaporation/condensation}

\begin{figure*}[t]
\begin{center}
  \includegraphics[width=12cm]{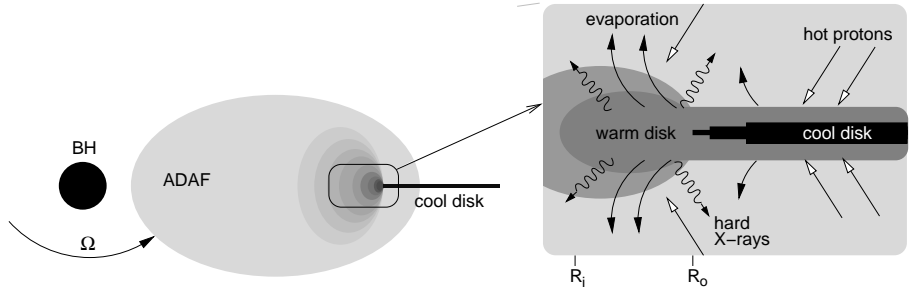}\caption{Evaporation from the
    warm disk: a cool disk (black, $\sim 1$ keV) partly extends into
    an ISAF and is exposed to the hot virialized protons from the hot
    torus (open arrows, light grey). Above and below the cool disk a
    heated surface layer is produced ($\sim 80$ keV, dark grey) due to
    proton illumination.  At the inner edge at $R_0$ the surface
    density of the cool disk gets small and a warm disk develops
    ($\sim 300-500$ keV, grey).  Net mass loss into the ISAF
    evaporates the warm region between $R_\mm{i}<R<R_0$ (filled
    arrows).  Due to the high temperatures in the warm disk this
    region is a source for hard X-rays.}
  \label{fig:evapgeo}
\end{center}
\end{figure*}

In the above we have started with the assumption that an ISAF and a
cool disk coexist, and found the conditions under which the disk can
feed mass into the ISAF. We have done this by considering the
conditions at each distance from the hole separately. To turn the
ingredients into a consistent picture, we have to consider the mass
flux through the system, so that conditions as a function of distances
from the hole are connected, in the way sketched in
Fig.~\ref{fig:evapgeo}. At large distance, we have a standard cool
disk. Closer in, an ISAF surrounds it and condenses onto it,
producing what we have called here the warm, proton-illuminated layer.
Even further in, the vertical optical depth of the disk becomes too
low to sustain cooling by brems losses, and the whole disk transforms
to a `warm disk' state (paper II) of nearly uniform temperature. The
upper layers of this warm disk evaporate to feed the ISAF assumed at
the outset, and at  the inner edge of the disk, at some radius $R_\mm{i}$, the
entire mass flux through the disk has evaporated into the ISAF. We now
investigate what the conditions are for such a radial structure to be
possible in a steady state.

The disk, whether in a cool ($\lesssim 1$ keV) or warm ($\sim 300$ keV)
state, is still very cold compared to the local virial temperature, so
that the standard thin disk approximation is valid.  The difference
with a standard steady disk is that the mass flux is now a function of
distance, due to the condensation and evaporation from/to the ISAF. We
first consider the modifications to the $\alpha$-disk diffusion
equation that this causes.

The surface mass density of the disk material is 

\begin{equation}
  \Sigma = \int_{-\infty}^{\infty} \rho \,\rmd z  \approx 2\rho_0\,H\; ,
  \label{eq:sigma}
\end{equation}
where $\rho_0$, $H$ denote the density at the midplane and the scale
height of the cool disk, respectively.

The change of the mass accretion rate with radius through the cool
disk due to evaporation can be expressed by

\begin{equation}
  \frac{\partial\dot M }{\partial R}= 4 \pi R \,\dot\mc{M}\; ,
  \label{eq:dmdotdr}
\end{equation}
where we include an additional factor 2 to account for the mass
loss on both sides of the disk. Integrating this equation yields the
mass loss (in g sec$^{-1}$) due to evaporation in a ring between $R_\mm{i}$
and $R$,

\begin{equation}
  \dot M_\mm{ev} = 4\pi \int_{R_\mm{i}}^{R} R \dot \mc{M} \rmd R\;.
  \label{eq:mev}
\end{equation}

The conservation of mass in a cool disk including the evaporation term
[Eq. (\ref{eq:dmdotdr})] can be expressed by

\begin{equation}
  \frac{\partial}{\partial t}(R\Sigma)+\frac{\partial}{\partial R}
  (R\Sigma v_\mm{R}) + R \dot\mathcal{M} = 0 \; ,
  \label{eq:cont}
\end{equation}
where $v_\mm{R}$ is the radial drift velocity of the disk material.

As in the standard derivation, the thin disk diffusion equation
follows from the angular momentum equation, which now includes a
term for the angular momentum carried with the
evaporating/condensing material. For the present purpose, it is
sufficient to assume that condensing and evaporating material just
has the same specific angular momentum, $\Omega_{\rm K}R^2= $ as the
disk. 

The equation for the angular momentum balance in an evaporating disk
then is

\begin{eqnarray}
  \frac{\partial}{\partial t}(R \Sigma \Omega R^2)+
  \frac{\partial}{\partial R}(R \Sigma v_\mm{R} \Omega R^2) +
  R\dot\mc{M} \, \Omega R^2 \nonumber  & = & \nonumber \\
  \frac{\partial}{\partial R}(S R^3 \frac{\partial \Omega}{\partial R})\; ,
\label{eq:angmom}
\end{eqnarray}
where

\begin{equation}
  S=\int_{-\infty}^{\infty} \rho\,\nu\,\rmd z \approx \Sigma\nu\;.
  \label{eq:S}
\end{equation}

We use the usual $\alpha$ prescription for the viscosity
\citep{sunyaev73}, 

\begin{equation}
  \nu = \alpha \frac{c_s^2}{\Omega}\;.
\label{eq:nu}
\end{equation}
The approximate equality in Eq.~(\ref{eq:S}) holds if $\nu$ can be
considered independent of the geometrical height $z$.

\subsection{Steady state}

Multiplying the continuity equation [Eq.~(\ref{eq:cont})] with $R^2
\Omega$ yields after subtraction from Eq.~(\ref{eq:angmom}) an
expression for the mass flux $\dot M$ in the cool disk in the
stationary case ($\partial/\partial t = 0$):

\begin{equation}
  \dot M = 6\,\pi R^{1/2} \frac{\partial}{\partial R}(\nu \Sigma R^{1/2})\;.
  \label{eq:mdot}
\end{equation}

The thin disk equations, Eqs.~(\ref{eq:cont},\ref{eq:angmom}) can be
  used in their time dependent form. For example, the evaporating
  inner regions could expand inward towards the last stable orbit, or
  outwards. As a first application, however, we are interested in
  steady-state conditions.  In the stationary case we can use
  Eq.~(\ref{eq:mdot}) to replace $\dot M$ in Eq.~(\ref{eq:dmdotdr}).
  Integrating this expression yields

\begin{equation}
  R^{1/2} \frac{\partial}{\partial R}(\nu \Sigma
  R^{1/2})= \frac{2}{3}\int_{R_\mm{i}}^{R} R \dot \mc{M} \rmd R + C_1\;.
  \label{eq:mdot6pi}
\end{equation}

We integrate one more time and obtain the surface density distribution:

\begin{equation}
  \nu \Sigma = \frac{2}{3 R^{1/2}} \int_{R_\mm{i}}^R \frac{\rmd R}{R^{1/2}}
    \int_{R_\mm{i}}^{R} R\,\dot\mc{M} \rmd R + 2 C_1 + \frac{C_2}{R^{1/2}}\;,
    \label{eq:nusig1}
\end{equation}
where $C_1,C_2$ are integration constants. Let the total mass loss due
to evaporation from $R_\mm{i}$ to infinity be $\dot M_\mm{ev,\infty}$ 
(cf. eq. \ref{eq:mev}). With Eq~.(\ref{eq:mdot},\ref{eq:mdot6pi}) we find the
integration constant $C_1$,

\begin{equation}
  C_1 = \frac{1}{6 \pi}(\dot M - \dot M_{\mm{ev},\infty})\;.
  \label{eq:c1}
\end{equation}

As in the standard treatment, the second integration constant is fixed
by considering the conditions at the inner edge of the disk. This
could be the last stable orbit if the disk extends all the way down to
the hole, but more interesting is the case when the {\it entire mass
flux has evaporated} into an ISAF {\it before} reaching the hole. Thus
we now assume that a steady state exists with the inner edge at some as
yet unspecified distance $R_{\rm i}$. The requirement of a steady
state will then impose a condition on the parameters of the system
that has to be satisfied. This will turn out to be a condition on the
accretion rate. At the end of the calculation, we thus
obtain a relation between the mass accretion rate and the position
of the inner edge.

Thus we set $\nu \Sigma_i=0$ at the inner edge of the disk, as in
standard accretion theory.  With
Eq.~(\ref{eq:mev},\ref{eq:nusig1},\ref{eq:c1}) and $R=R_\mm{i}$ we get
an expression for the second integration constant $C_2$,

\begin{equation}
  C_2 = -\frac{R_\mm{i}^{1/2}}{3\pi}(\dot M -\dot M_{\mm{ev},\infty})\;.
  \label{eq:c2}
\end{equation}

With the previous results the surface density distribution as a
function of radius in a disk with evaporation losses is now 

\begin{eqnarray}
  \nu\Sigma  =  \frac{1}{6\pi\,R^{1/2}}\int_{R_\mm{i}}^{R}\frac{\rmd
R}{R^{1/2}}
  \dot M_\mm{ev}(R)\; + \\ \nonumber 
  \frac{1}{3\pi}\,[\dot M - \dot
  M_{\mm{ev},\infty}]\,[\, 1 - (R_\mm{i}/R)^{1/2}]  \; .
  \label{eq:nusigma}
\end{eqnarray}

From this expression the classical formula for the surface density in
a cool disk can be recovered if one sets $\dot M_\mm{ev}=0$, i.e. when
no evaporation takes place.

This expression for the surface density, though strictly derived for
steady conditions, is still approximately valid if the position of the
inner edge changes slowly. We are interested in a true stationary
case, however. In this case the accretion rate in the disk will be
exactly equal to the total mass loss due to evaporation in the disk,
$\dot M = \dot M_{\mm{ev},\infty}$, or in other words all matter
drifting inward through the cool disk has evaporated when $R_\mm{i}$ 
is reached. Eq.~(36) then simplifies to

\begin{equation}
  \nu \Sigma = \frac{1}{6\pi\,R^{1/2}}\int_{R_\mm{i}}^{R} \frac{\rmd
  R}{R^{1/2}} \dot M_\mm{ev}(R).
  \label{eq:nusigsim}
\end{equation}

\subsection{Radial extent of the warm disk}

We can now estimate the distance over which the process of evaporation
into the ISAF takes place. Let $R_0$ be the innermost radius where the
cool disk component exists. Inside this, there is only a warm disk
(see Fig. \ref{fig:evapgeo}). Evaporation takes place both from the
warm layer on top of the cool disk at $R>R_0$ and from the warm disk
region $R_{\rm i}<R<R_0$, but the warm disk region is expected to
contribute most, since its temperature is significantly higher. Thus
we equate, for the present approximate purpose, the evaporating region
with the warm disk region. Assume that the relative extent
$\delta_0=R_0/R_{\rm i}-1$ of the warm disk is small. The evaporation rate
Eq.~(\ref{eq:evap}), with $\Omega\approx \Omega_{\rm i}$, depends only
on temperature. The temperature of the warm disk is relatively fixed
(paper II), so we can set $\dot \mc{M}$ constant as well.
Eq.~(\ref{eq:nusigma}) for the surface density as a function of
distance from the inner edge is then

\begin{equation}
  \nu \Sigma = \frac{2}{3}\int_{R_\mm{i}}^{R}\rmd
  R\int_{R_\mm{i}}^{R}\dot\mc{M}\,\rmd R \approx
  \frac{1}{3} \dot\mc{M}\,(R-R_\mm{i})^2 ,
  \label{eq:nusig2}
\end{equation}
or
\begin{equation}
  \Sigma = \frac{1}{3} \mc{A}\;\delta_0^2\;,
  \label{eq:sigma1}
\end{equation}
where

\begin{equation}
  \mc{A} = \frac{\Omega}{\alpha\,c_s^2}\,\dot\mc{M}\,R_\mm{i}^2 \; .
  \label{eq:a1}
\end{equation}

In the warm state, little net condensation takes place (as we have seen in
Sect.~\ref{sec:visvscon}) and $\dot\mc{M}$ is dominated by
the evaporation losses. Using Eq.~(\ref{eq:evap}) for $\dot\mc{M}$
we then find

\begin{equation}
  \mc{A} =
\alpha\,\theta\,f\,\frac{3}{16\sqrt{2\pi}\,\ln\Lambda}\,\frac{m_\mm{p}}{r_0^2}
  \left(\frac{m_\mm{p}}{m_\mm{e}}\right)^{3/2}
  \frac{R_\mm{S}}{R_\mm{i}} \; ,
  \label{eq:a2}
\end{equation}
where $r_0$ is the classical electron radius and $\theta=k
T_\mm{e}/m_\mm{e}c^2$. This result depends on the temperature of the warm
disk, the viscosity parameter and the distance from the hole. 

We can now make an estimate of the relative extent $\delta_0$ of the warm
disk. $R_0$ is the maximum radius where the warm disk can
exist (at larger surface density, it develops a cool disk component).
The optical depth at $R_0$ is thus given by Eq.~(\ref{eq:taum}). Computing
the optical depth from Eq.~(\ref{eq:sigma1}) and equating this to $\tau_{\rm m}$ 
we have

\begin{equation}
  \kappa \,\frac{\Omega \dot \mc{M} R_\mm{i}^2}{3\alpha\,c_s^2}\,\delta_0^2
  \approx 3\left(\frac{\dot M_\mm{I}}{0.1 \cdot \dot
M_\mm{Edd}}\right)^{1/2}\;,
  \label{eq:tau}
\end{equation}
which shows that $\delta_0$ is a function of the warm disk temperature, the
viscosity,
the relative distance from the hole and the accretion rate. Numerically, we have

\begin{eqnarray}
\delta_0 \approx
     0.1  \left(\frac{1}{\alpha\,\theta \,f}\right)^\frac{1}{2}
  \,\left(\frac{R_\mm{i}}{R_\mm{S}}\right)^\frac{1}{2}
\,\left(\frac{\dot M_\mm{I}}{\dot M_\mm{Edd}}\right)^\frac{1}{4}\;,
      \label{eq:x0}
\end{eqnarray}
where we have set $\ln\Lambda\approx 20$.  With typical values, $\theta\sim
O(1)$,
$R_\mm{i}/R_{\rm S}\sim 8$, $\dot M_\mm{I}/\dot M_\mm{Edd}=0.1$,
$f=0.88$ and $\alpha=0.3$ this yields $\delta_0 \approx 0.3$.  We
conclude that the extent of the warm disk is in an `interesting'
range. It is smaller than the distance to the hole, but still large
enough that the width of the warm disk is large compared to its
thickness, so our thin-disk treatment of the evaporation region is
justified.

\subsection{Conditions for a steady mass flux to exist}
We now estimate the radial distance of the transition radius from the
hole as a function of the mass accretion rate $\dot M$ in the system.
At the distance $R_\mm{i}$ all mass accreted in the cool disk is
evaporated, $\dot M_\mm{ev} = \dot M$.  The accretion rate $\dot M$ of
the system is now also equal to the accretion rate $\dot M_\mm{I}$ in
the ISAF and we can write

\begin{equation}
\dot M_\mm{ev} \approx 2 \cdot 2 \pi \,R_\mm{i} \Delta R_\mm{i}\,
\dot\mc{M} =
        4\pi\,R_\mm{i}^2\,\delta_0\,\dot\mc{M} = \dot M_\mm{I}
\label{eq:evest}
\end{equation}

We use Eqs.~(\ref{eq:evap},\ref{eq:x0}) and find for $R_\mm{i}$ (in
units of the Schwarzschild radius)

\begin{equation}
\frac{R_\mm{i}}{R_\mm{S}} \approx \;7 \times
\frac{1}{\alpha^{3/2}\,\theta^{3/2}\,f^{1/2}}
\left(\frac{\dot M_\mm{I}}{\dot M_\mm{Edd}}\right)^\frac{3}{4}
\label{eq:rivonmdot}
\end{equation}
Again with the same values as above,
$\theta\sim O(1)$, $\dot M_\mm{I}/\dot M_\mm{Edd}=0.1$, $f=0.88$ and
$\alpha=0.3$ this yields $R_\mm{i}/R_\mm{S}\approx 8$. We conclude
that, using plausible values, the estimated transition radius is in
the region where we expect our model to work.

\section{Discussion and conclusions}

Our model describes a situation which naturally develops if a cold
standard Shakura--Sunyaev disk is truncated within a hot, optically
thin flow (ISAF) in the inner regions of the accreting system. We have
proposed a SSD--ISAF transition based on few, well--known physical
processes: Spitzer's theory of the energy exchange in
a fully ionized plasma, and standard viscous heating due to friction in
the accretion disk.  We have shown that the transition of the cool disk
material into an ISAF is the logical and inevitable
consequence of these basic interactions. The process involves two
steps:

(i) At the inner edge of the disk the surface density of the cool
disk gets low.  Virial protons penetrating from the ISAF heat the cool
disk electrons. The electron temperature is limited ($T_\mm{e}\sim 80$
keV) since they can radiate their energy efficiently via bremsstrahlung and
Comptonization. Once the disk is too thin, proton heating overcomes
the radiative losses everywhere in the disk. The disk heats up,
expands and radiative losses become even more inefficient. With
increasing temperature pair production sets in, but also the proton
heating gets less efficient at higher temperature. Finally a new
equilibrium for the thin disk is found at several 100 keV. We label
this the `warm state', since its temperature is intermediate between
that of the cool disk and the virial temperature.  An important aspect in 
this process is that protons in the disk are outside of the main energy 
channel. The main energy is transferred from the external ISAF protons 
to the disk electrons, which loose this energy via radiation. For the 
formation of the warm disk state the internal viscous heating of the 
protons is completely unimportant, but not so in the second step of
the process:

(ii) At the temperatures of the warm state the disk protons and
electrons are not coupled very tightly any more, as the timescale for
establishing thermal equilibrium is not short compared to the thermal
timescale. Now the minor energy channel due to viscous heating in the
warm disk becomes important for the energy budget of the protons, because
the viscously released energy can not be exchanged very efficiently
with the ambient electrons. The upper part of the warm disk, where 
the densities are lowest and the Coulomb exchange time scales longest, is 
subject to a thermal instability. The size of the unstable region 
depends mainly on the viscosity parameter and the temperature of the warm
disk. The higher the temperatures and $\alpha$, the deeper is the
unstable region. The protons there are heated to their local virial
temperature, and become part of the ISAF: the warm state evaporates.
In the warm state no effective mass condensation takes place, since it 
at the same time becomes optically thin for penetrating hot
protons.  The mass evaporation due to the viscous instability of the
warm state is therefore the main process which determines the
mass budget in this region.

On the basis of this picture, we have also looked at the radial
structure of an accretion disk in which mass exchange with an ISAF
takes place. Using the evaporation and condensation rates derived, we
have investigated the conditions under which accretion is possible in
such a way that the entire accretion flow eventually evaporates. In
this case a steady state is possible with an inner edge to the disk at
some finite radius $R_{\rm i}$ outside the last stable orbit. We find
that such a steady state is indeed possible for plausible values for
the accretion rate and viscosity parameter. The steady state condition
determines a relation between the accretion rate and the value of
$R_{\rm i}$. Also, it determines the width $\delta=\Delta R/R_{\rm i}$
of the warm, evaporating disk region; we find $\delta\approx 0.3$ for
$\alpha$ of order 0.3 and accretion at a tenth of Eddington.

A potentially important factor which we have not been able to include
in our picture of the warm disk region is the cooling effect of soft
photons from the cool disk extending just outside the warm disk. If
such photons can enter the warm region, they will cause a cooling
by inverse Compton scattering on the warm electrons.
Since both the cool and warm disk are quite thin ($H/R\ll 1$), the radial 
optical depth of the warm disk is large, and the angle subtended by
the cool disk as seen from the interior of the warm disk is small (to
visualize this, see Fig. 5). It thus seems likely that the effect of
such cooling on the energy balance in the warm disk can be small, but this
point requires closer scrutiny.

The extent of the region (in distance from the hole) where the
physical conditions assumed here apply is limited.  At large radii
the proton temperature in the optically thin, hot region decreases
($\propto R^{-1}$). At lower proton temperatures the proton
penetration depth into the cool disk also gets smaller. This limits the
region where a warm state can be produced. Without the warm state our
mechanism will probably not work efficiently enough to transfer all
material from a cool SSD into material from an ISAF.  Therefore we do
not expect our mechanism to work at large radii from the black hole.
But a combination of the coronal evaporation flow model as suggested
by \cite{meyer00} and the two stage model proposed here could in
principle cover a large range in the radial direction for the
SSD--ISAF transition to occur.

\begin{acknowledgements}
  This work was done in the research network ``Accretion onto black
  holes, compact stars and proto stars'' funded by the European
  Commission under contract number ERBFMRX-CT98-0195.
\end{acknowledgements}

\bibliographystyle{apj}
\bibliography{aamnem99,pub}

\end{document}